# Thermally assisted skyrmions creation in Pt/Co/Ta multilayer films

Senfu Zhang[1#], Junwei Zhang[1#, 2], Yan Wen[1], Eugene M. Chudnovsky[3] and Xixiang Zhang[1*]

[1]*Physical Science and Engineering Division (PSE), King Abdullah University of Science and Technology (KAUST), Thuwal 23955-6900, Saudi Arabia*

[2]*Key Laboratory for Magnetism and Magnetic Materials of Ministry of Education, Lanzhou University, Lanzhou, 730000, People's Republic of China*

[3]*Physics Department, Lehman College and Graduate School, The City University of New York, 250 Bedford Park Boulevard West, Bronx, New York 10468-1589, USA*

Néel-type magnetic skyrmions in multilayer films have attracted significant amount of attention recently for their stability at room temperature and capability of motion driven by a low-density electrical current, which can be potentially applied to spintronic devices. However, the thermal effect on the formation of the skyrmions and their behavior has rarely been studied. Here, we report a study on the creation of skyrmions in [Pt/Co/Ta]$_{10}$ multilayer samples at different temperatures using an in-situ Lorentz transmission electron microscopy. By imaging the magnetization reversal process from positive (negative) saturation to negative (positive) saturation, we found that the skyrmions can be created by nucleation from ferromagnetic saturation state and by breaking the labyrinth domains under certain external fields. By tuning the external fields, a maximum density of skyrmions was reached at different temperatures. The key finding is that the creation of the skyrmions in the multilayers depends critically on the temperature and thermal history.

[#] Equally contributed to this work.

[*] E-mail address: xixiang.zhang@kaust.edu.sa (Xixiang Zhang)

Magnetic skyrmions are stable swirling spin textures, which have attracted tremendous interests recently due to their topological protected property with very small size and novel current induced dynamics[1-5]. These extraordinary properties make skyrmions have potential applications in next generation spintronic devices[6-12]. Skyrmions were widely observed in both bulk materials[13-24] and multilayer films[5, 25-33]. Most of skyrmions in bulk materials are of Bloch-type and can only survive at relatively low temperatures[3, 4, 13-24]. To create room temperature skyrmions suitable for industrial applications, heavy metal (I) / ferromagnet / heavy metal (II) (or oxide) hetero-structured multilayer films were developed, in which Dzyaloshinskii-Moriya interaction (DMI), originating from the strong spin-orbit coupling of interfacial atoms neighboring the magnetic layer, is the key for the formation of skyrmions[5, 25-33]. In the herterostrycured magnetic multilayers, DMIs at the two interfaces between the ferromagnetic layer and two dfferent heavy metal layers are very different. Morever, for the ferromagneic layer, the lack of the inversion symmetry will enhace the total DMI acting on the ferromagnetic layer greatly[4, 27]. Néel-type skyrmions have been observed in wide range of thin-film stacks by the interplay of DMI, dipole interaction, perpendicular magnetic anisotropy, exchange interaction and the external magnetic field[5, 25-33]. The mechanism of formation and how to control the density of the skyrmions in those films have also been studied[8, 29, 31, 34]. To increase the density of the skyrmions, in general, one can decrease the perpendicular anisotropy $K_u$ by tuning the thickness of the ferromagnetic [30, 33, 35] or to increase the DMI by choosing proper heavy metals[5, 27, 33]. It has also found that for some multilayers, the electrical current[36], in-plane external magnetic field and the film disorder[37] will also facilitate the creation of skyrmions. We recently discovered that skyrmions lattice could also been directly written by the magnetic force microscopy (MFM) tips at zero magnetic field[38].

To increase stability of the skyrmions against the thermal fluctuation at high temperatures, a multilayered structure, composed of a number of tri-layer unit of heavy metal (I) / ferromagnet / heavy metal (II) (or oxide), is a common strategy. The columnar-shaped skyrmions created in the thick multilayers have a much larger volume than that in a thin tri-layers, consequently are much more stable at room temperature. [4, 5, 27, 29, 31-33, 38] However, it is still not well known whether skyrmions survive at a higher temperature and how the temperature affect the creation of skyrmions. In this work, we investigated the magnetization reversal behaviors in [Pt/Co/Ta]$_{10}$ multilayer samples at different temperatures using in-situ, Lorentz transmission electron microscopy (L-TEM) and found that the skyrmion density increases substantially with increasing temperature. Moreover, high density skyrmions were also created at high temperature in the sample with strong PMA where no skyrmions could be created at room temperature.

The multilayer stacks of Ta(5 nm)/[Pt(3 nm)/Co(2 nm)/Ta(2nm)]$_{10}$ were deposited by DC magnetron sputtering on both the Si substrates and Si$_3$N$_4$ membranes at room temperature. Magnetic properties of the multilayers were characterized using a SQUID magnetometer in a very wide temperature range from 5K to 600K. Since L-TEM is one of the most direct methods to observe the skyrmions with high spatial resolution[32, 37], we performed L-TEM observations of the magnetization reversal at different temperatures using both the cooling in-situ holder and the heating in-situ holder.

To understand the basic properties of the multilayers, we measured the hysteresis loops at different temperatures with magnetic field applied parallel and perpendicular to the film planes. Figure 1(a) shows the normalized out-of-plane and in-plane hysteresis loops measured at 300 K. The behavior of the loops indicates clearly that the sample has a perpendicular magnetic anisotropy (PMA). The effective perpendicular anisotropic field (or in-plane saturation field) $H_k$ is about 2.9 kOe.

The reversal behaviors of the magnetization in these multilayers have also studied with in-situ L-TEM at different temperatures. Unless stated otherwise, all the L-TEM images were taken at a defocus of 8.64 mm and a tilt angle of 30$^\circ$. Figure1 (b)-(f) shows five representative L-TEM images taken during sweeping the magnetic field from the positive saturation field to the negative saturation at 300 K. It is clearly seen that when the magnetic field is decreased from the positive saturation magnetic field to a 560 Oe field (Fig. 1(b)), some isolated skyrmions as well as snake-like strip domains emerged from the ferromagnetic state. This process will be designated as "nucleating process". As the field is continually decreased to 520 Oe, all the skyrmions and the snake-like domains extended to much longer snake-like domains with an preferred orientation[37] as shown in Fig. 1(c). With the field being further decreased, the snake-like domains become longer and longer. At zero field, the density of snake-like domains achieves a maximum over the entire film and a labyrinth domain structures forms as shown in Fig. 1(d). As the field is increased in the opposite direction to a negative field (-580 Oe), the labyrinth domain structures fractures into small pieces and a mixed state of skyrmions and snake-like domains appears as shown in Fig. 1(e). More and more skyrmions are created and eventually, the skyrmion density reaches the maximum when the field is increased to -840 Oe (Fig. 1(f)). We will then call the creation process of the skyrmions through fracturing the snake-like domains as "breaking process". With increasing the field further, the skyrmions annihilates gradually and the sample finally reaches the saturated, ferromagnetic state in the opposite direction and the magnetization reversal in the whole sample have completed.

Previous literatures[33] have shown that the density of skyrmions in a material increases with increasing the material parameter $\kappa$, where $\kappa = \pi D / 4\sqrt{AK_{eff}}$, $D$ is the DMI constant, $A$ is the exchange stiffness and $K_{eff}$ is the effective perpendicular anisotropy. Since the values of $D$ and $A$ are intrinsic properties of the materials and do not change with temperature, an effective method to increase the parameter $\kappa$ is to decrease $K_{eff}$[39-42]. To achieve this goal, one can increase the sample temperature, because the anisotropy constant of magnetic materials decreases with increasing temperature. As it is well known, the anisotropy field/constant can be calculated from the hysteresis loops. To obtain the temperature dependence of magnetic anisotropy constant, the hysteresis loops of the multilayers were measured at different temperatures (see details in supplementary Fig. 1). Figure 1g and 1f show the temperature dependence of the saturation magnetization $M_s$ and the effective perpendicular anisotropy field $H_k$. It is evident that both $M_s$ and $H_k$ decrease monotonically with increasing temperature. As $K_{eff}$ is a product of $M_s$ and $H_k$, i.e. $K_{eff} = \frac{\mu_0 H_k M_s}{2}$, $K_{eff}$ should have a stronger temperature dependence and decreases faster than both $M_s$ and $H_k$ as the temperature increases.

To verify the above argument that skyrmion density increases with increasing temperature, magnetization reversal were further studied using L-TEM in the temperatures range from 97 K to 513 K. Figure 2 (a) shows the L-TEM images taken at 97 K, 453 K and 513 K and when the density of the skyrmion reach the maximum in the nucleating and breaking process respectively. It is clearly seen that at low temperatures, nucleation of skyrmions from a ferromagnetic state is very difficult, i.e. only a few skyrmions nucleated at 97 K. However, as reported in Ref[37], much more skyrmions form in the breaking process, as the applied magnetic field is increased in the opposite direction from zero. At 97 K, the density of skyrmion reaches a maximum at field of -980 Oe. Interestingly, with increasing the temperature, we found that the density of skyrmions in both the nucleating and the breaking processes increases dramatically. Figure 2(b) shows the temperature dependence of the maximum densities of the skyrmion, which strongly indicates that the thermal fluctuation plays the most critical role in the formation of skyrmions. Moreover, we found that the absolute values of the magnetic fields at which the maximum density of the skyrmions occurs decreases from 710 Oe to 530 Oe in the nucleating process and from 980 Oe to 580 Oe in the breaking process, as the temperature increases from 97 K to 513 K. In Fig. 2(c), we plot the temperature dependence of the magnetic fields at which the density of skyrmions reaches the maximum in the breaking process and at which the sample reaches the saturated ferromagnetic state. It shows that both fields depend on temperature very weakly in the low temperatures range of 97 K to 253K and that the strength of both fields decreases significantly in the high temperature range of 300 K to 513 K. We can then conclude that thermal fluctuation indeed facilitates the creation of skyrmions under certain circumstance.

To gain a deeper understanding of the thermal effect on the formation of skyrmions, we studied the evolution of the magnetic structures with temperature under a constant external field using the L-TEM. To this end, a strong and positive magnetic field was first applied to saturate the sample and then the field was decreased to 580 Oe at 300 K. The L-TEM image shows that a few isolated skyrmions and snake-like structures emerged from the ferromagnetic state as shown in Fig. 3(a). By keeping other parameters unchanged, the temperature was increased to 513 K gradually. The L-TEM images were then taken at different temperatures during the temperature sweeping (Fig. 3(a)). We found that the snake-like domains broke into skyrmions with increasing temperature and that more and more new skyrmions nucleated gradually from the ferromagnetic state, as represented by the images taken at 443 K and 513 K. Note that, in order to avoid annealing effect that may change the properties of the sample significantly, the highest temperature used here is 513 K. Interestingly, we found that the skyrmions state observed at 513 K kept unchanged even after the temperature was decreased back to 300 K. To examine whether the property of the sample changed after above heating process, we repeated the same experiments at 300 K and observed the same (very similar) reversal behavior as shown in Fig. 1(b-f). We can then confirm that cycling temperature between 300 K and 513 K will not significantly alter the properties of the sample. We repeated the experiment with the same temperature cycle but under a smaller magnetic field; that is, a strong and positive field was applied to saturate the sample, then the field was decreased to 520 Oe at 300 K. Fig. 3(b) shows the L-TEM images taken during the thermal cycles, 300 K→513K→300K. We found that after the field was deceased to a smaller field (520 Oe) at 300 K, more snake-like structures appeared in the sample and that a high density skyrmions state was achieved in the film at 513 K. This density is much higher than that obtained at room temperature under any condition without raising the temperature, for example, the results shown in Fig. 1. Another interesting finding in above experiments is that the high density, individual skyrmions can survive even at a field of 520 Oe at 300 K after cooling down from 513 K. This field is significantly smaller than 600 Oe that is required to stabilize isolated skyrmion in this sample at room temperature (without raising the temperature). We also find that a few skyrmions have a tendency to extend to snake-like structure as marked by the red dotted ellipses (Fig. 3(b)). This can be understood as following. For an isolated skyrmion, it will extend to a longer structure when the magnetic field is lower than the critical field. While for a skyrmions lattice with very high density, the repulsive forces from the neighboring skyrmions enhance the stability of the skyrmion[38].

We have presented above the temperature effect in the skyrmion nucleating process, as shown in Fig. 3 (a) and (b). We now turn to the temperature effect on the formation of skyrmions in the breaking process. We saturate the film again with a positive magnetic field at 300 K. After that, the external field is gradually changed to negative field (-580 Oe), where the long labyrinth domains just started to break into small pieces as shown in Fig. 3 (c) (similar as the state shown in Fig. 1(e)).With increasing the temperature, all stripe domains transform to skyrmions gradually (see the image obtained at 408 K) and skyrmions lattice with a very high density forms at 513 K. Note that, the density of the skyrmions at 513 K is much higher than that obtained in the nucleating process at the same temperature, as shown in Fig. 3(a). This is because the creation of skyrmions is relatively easier in breaking process than in the nucleating process. Similarly, all the skyrmions kept stable during cooling the sample down to 300 K.

To investigate the temperature effect on the multilayers film with a stronger perpendicular magnetic anisotropy (PMA), we deposited a [Pt(3 nm)/Co(1.6 nm)/Ta(2 nm)]$_{10}$ multilayer sample. The hysteresis loops were measured in the temperature range of from 5 K to 850 K. Figure 4a shows the normalized hysteresis loops measured at room temperature (see the detailed

M-H loops at other temperatures in supplementary Fig. 2). Similarly, both $M_s$ and $H_k$ decrease with increasing temperature as shown in Fig. 4b. It is apparent that the Curie temperature of this sample should be a round 650 K, based on the behavior of the curves in Fig. 4b. More importantly, the value of $H_k$ in this [Pt(3 nm)/Co(1.6 nm)/Ta(2 nm)]$_{10}$ multilayer reaches as high as 7.2 kOe at room temperature, being much higher than that observed in [Pt(3 nm)/Co(2 nm)/Ta(2 nm)]$_{10}$ multilayer (Fig. 1b). Due to the strong perpendicular anisotropy, the domain width in this sample can be as large as about 810 nm at zero external magnetic fields and no skyrmions could be created at room temperature in neither nucleation process nor breaking process as expected[29, 30, 43] (see the detailed reversal behaviors in supplementary movie 1). We then heated the sample and changed the magnetic field from negative saturation field to positive saturation field, during which we observed the reversal of the magnetization (Note that, since the reversal behaviors during the field sweeping from negative saturation field to positive saturation field is same as that during sweeping field from positive saturation field to negative saturation field as expected for a symmetrical hysteresis loop, the breaking process takes place at positive field in this experiment). Figure 4(c) shows the L-TEM images of the sample at zero external field (first row) and the images when skyrmions density reaches the maximum at the temperature of 323 K, 593 K and 633 K by tuning the applied external magnetic field in breaking process (the second row). The domain width, at zero external magnetic field, decreases significantly with increasing temperature as observed in the first row of Fig. 4(c). To quantify our results, we measured the domain widths at different temperatures and plot it as function of temperature in Fig. 4 (d). With increasing temperature from 300 K to 633 K, the domain width decreased nearly linearly from about 810 nm to about 140 nm. Increasing the temperature further, the labyrinth domains thinned gradually and eventually faded out at 673 K as shown in the first row of Fig. 4(c). The evolution of images from 633 K to 673 K can be ascribed to the transition from the ferromagnetic state to the paramagnetic state as the temperature approached to the Curie temperature. Since there is no spontaneous magnetization in a paramagnetic state where the magnetic spins point randomly in the space, the contracts induced by magnetic domains in the Lorentz-TEM image disappear. At temperatures lower than 553 K, no skyrmions formed at any magnetic field as shown by the image taken at 323 K in Fig. 4(c) (See more information in Supplementary Fig. 3). We found that with increasing the temperature to decrease the domain width, more and more skyrmions formed gradually. We mainly focus on the skyrmions creation in the breaking process in the following study. When the temperature was increased to 593 K, some isolated skyrmions appear under a 400 Oe magnetic field, as observed in the second row of Fig. 4(c). At 633 K, a skyrmion lattice formed eventually with the external field of 410 Oe. In Fig. 4(d), we plot the skyrmions densities obtained at several elevated temperatures (the stars).

Although neither labyrinth domain structure nor skyrmions exist when the temperature was higher than 673 K (the Curie temperature), when the sample was cooled down in a constant, small magnetic field of 290 Oe (the field-cooling process) to 300 K, high density skyrmions gradually emerged and their L-TEM images became clearer and clearer with decreasing the temperature. The last image in the second row of Fig. 4(c) shows the skyrmion lattice formed after field-cooling process from 673 K to 300 K in a relative small magnetic field of 290 Oe. We then performed more field-cooling experiments, cooling the sample from different initial high temperatures to 300 K in the same magnetic field of 290 Oe. The density of the skyrmions observed at 300 K after field-cooling process with different initial temperature is then plotted in Fig. 4 (d) (the squares). The density increases with increasing the initial high temperature in the range of 650 K to 733 K, then saturated with value as high as 12 $\mu m^{-2}$. According above results, we can claim that the high density, uniformly distributed skyrmions lattice can be created much more easily through the field cooling process with a much smaller external magnetic field, particularly for high anisotropy materials. This finding should be of great importance for potential application.

In summary, the magnetization reversal behaviors in [Pt/Co(1.6 nm)/Ta]$_{10}$ and [Pt/Co(2.0 nm)/Ta]$_{10}$ multilayer samples were investigated at different temperatures by using the in-situ L-TEM. We show that thermal effect plays a critical role in the creation of skyrmions. We also show that high density, uniformly distributed skyrmions can be much more easily created through field cooling processes, which should be very important for practical application. A detailed, quantitative study of the thermal effect is undergoing.

# ACKNOWLEDGEMENTS

This publication is based on research supported by the King Abdullah University of Science and Technology (KAUST), Office of Sponsored Research (OSR) and under the award No. OSR-2016-CRG5-2977.

Figure 1

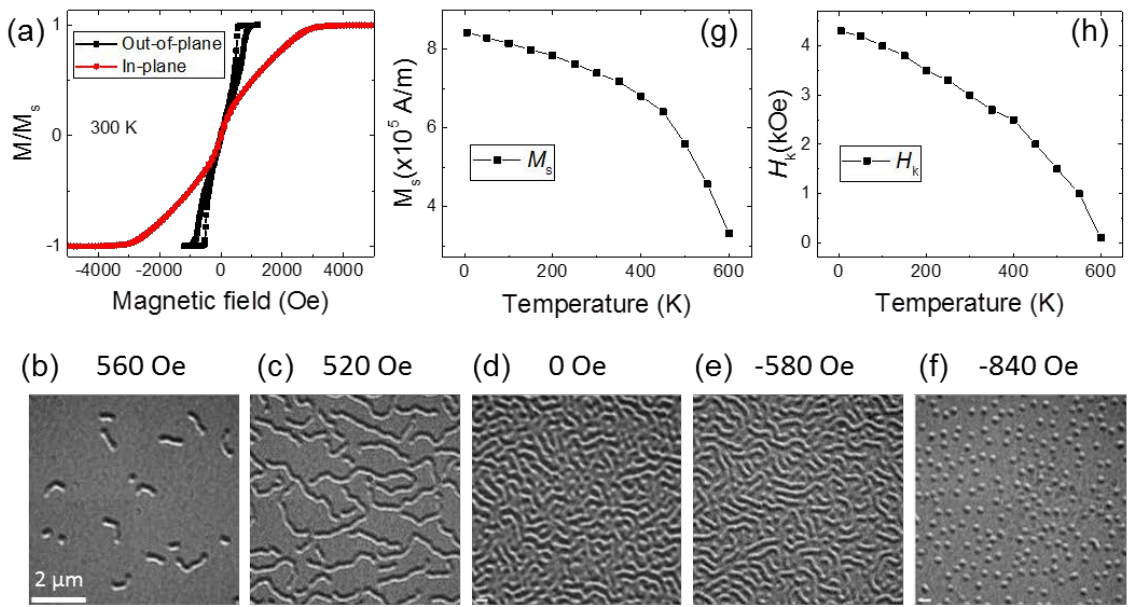

Figure 1. (a) Normalized out-of-plane and in-plane hysteresis loops of the [Pt(3 nm)/Co(2 nm)/Ta(2 nm)]$_{10}$ multilayer sample measured using SQUID magnetometer at 300 K. (b)-(f) In-situ L-TEM observation of the reversal behaviors taken during changing the magnetic field from positive saturation field to negative saturation at room temperature. The images were taken at the defocus of 8.64 mm and a tilt angle of 30°. (The field of -840 Oe correspond the maximum density of skyrmions at room temperature). (g-h) Temperature dependence of the (g) saturation magnetization $M_s$ and (h) effective anisotropy field $H_k$.

Figure 2

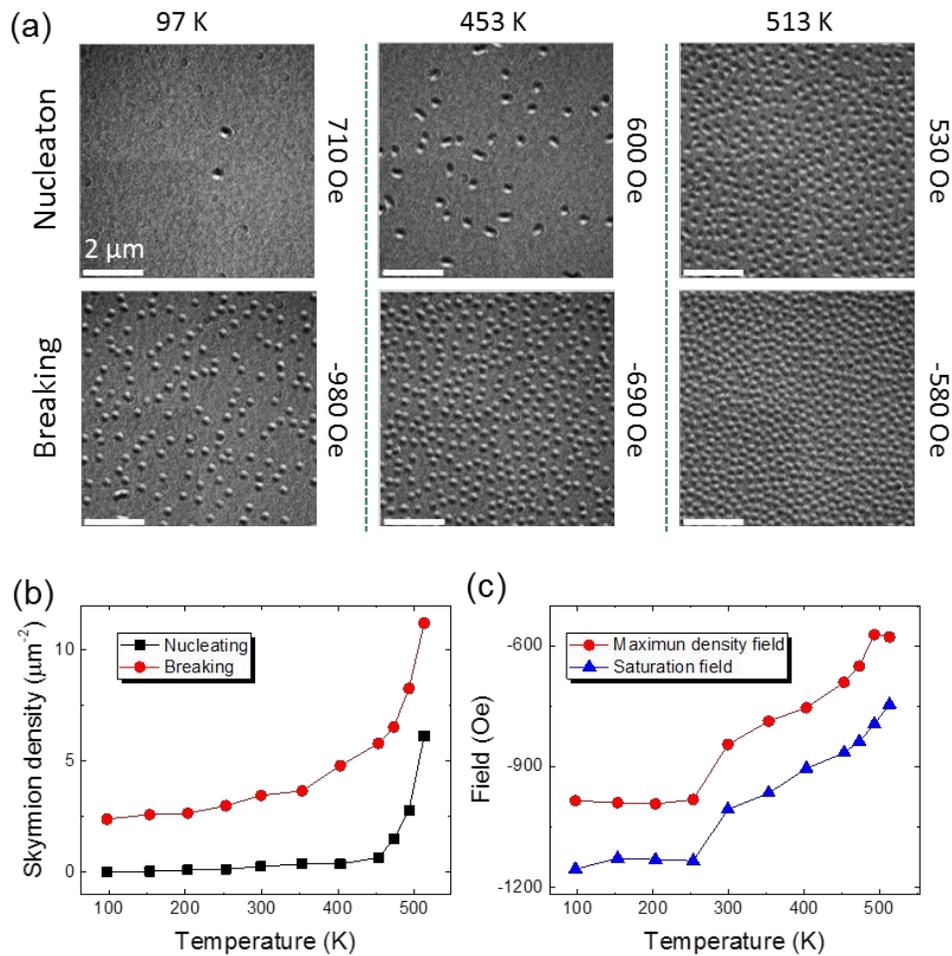

**Figure 2.** (a) L-TEM images when the skyrmion densities reach the maximum in the nucleating (demagnetizing process by reducing field from positive saturation) and domain breaking process (magnetizing process after passing through the coercive field) at 97 K, 453 K and 513 K, respectively. (b) The maximum skyrmion densities in the nucleating process and breaking process as functions of the temperature. (c) The field when the skyrmion density reaches the maximum in the breaking process and the saturation field as functions of the temperature.

Figure 3

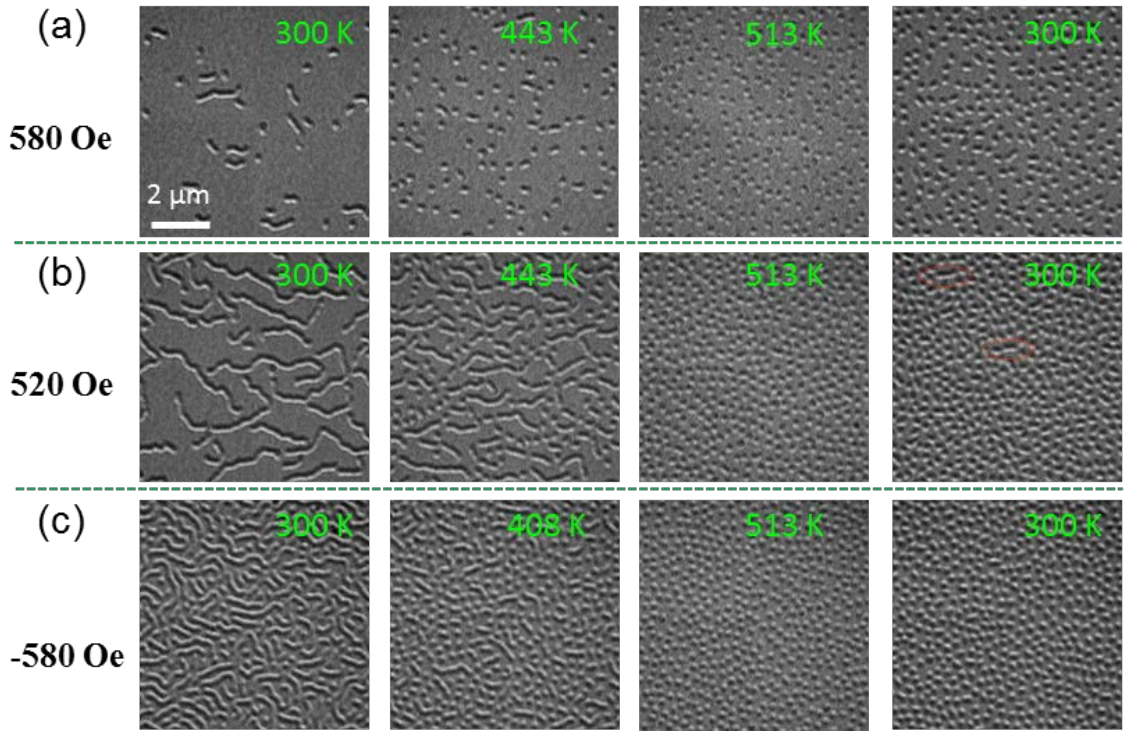

**Figure 3.** In-situ L-TEM observation of the skyrmions creation process by increasing the temperature from 300 K to 513 K and go back to room temperature (300 K) with a constant external field of (a) -580 Oe, (b) -520 Oe and (c) 580 Oe.

Figure 4

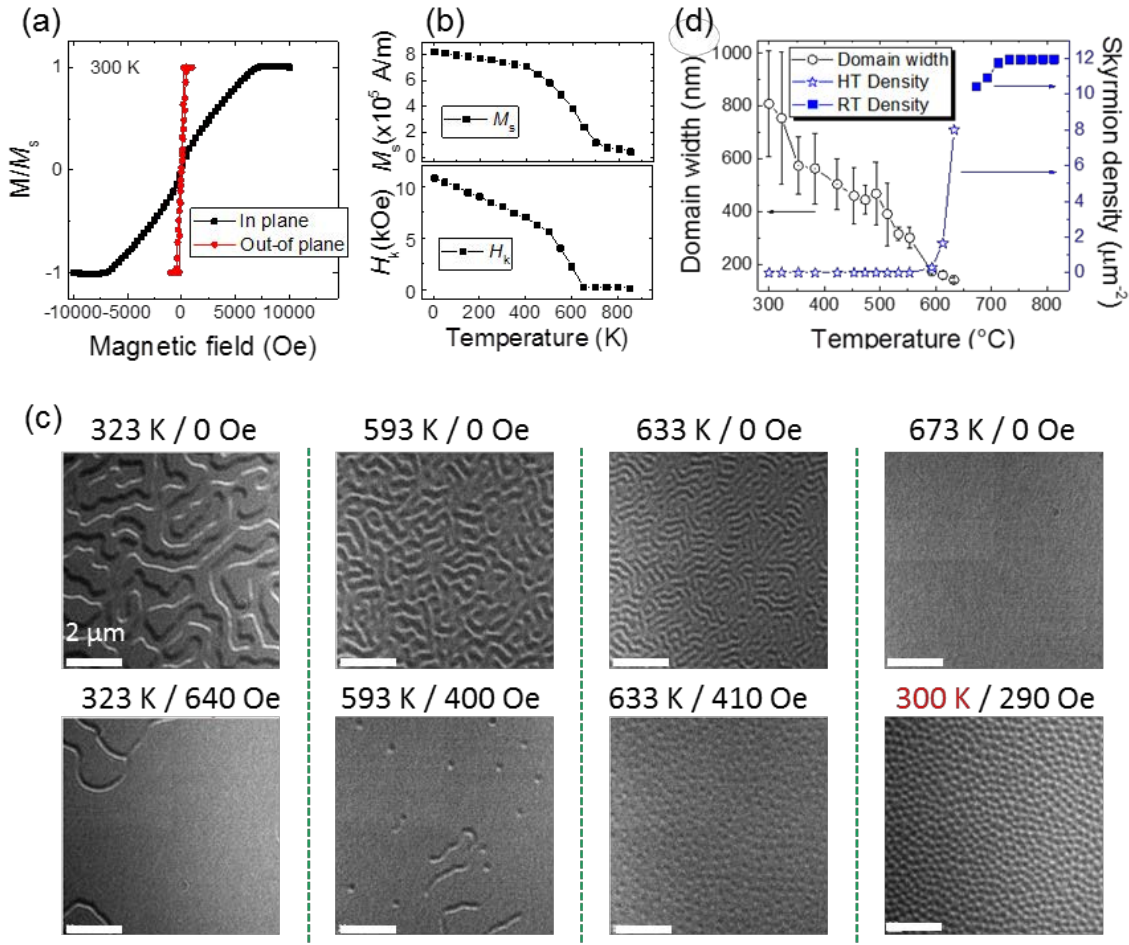

**Figure 4.** (a) Normalized out-of-plane and in-plane hysteresis loops of the [Pt(3 nm)/Co(1.6 nm)/Ta(2 nm)]$_{10}$ multilayer sample. (b) Temperature dependence of the saturation magnetization $M_s$ and the effective anisotropy field $H_k$. (c) L-TEM images of the sample at zero external fields and the corresponding images when skyrmions density reach the maximum in the breaking process at the temperature of 323 K, 593 K, 633 K and 673 K. (d) The strip domain width at zero external field and the maximum skyrmions density as a function of the (initial) temperature. The blue five-pointed star represents the skyrmion density observed at high temperature (HT). While the blue square represents the skyrmion density observed after decrease the temperature from initial high temperature to room temperature (RT) with a constant field of 290 Oe.

15